\documentclass[aps,prl,twocolumn,showpacs,superscriptaddress,tightenlines,floatfix]{revtex4}

\usepackage{amsmath}
\usepackage{amssymb}
\usepackage{graphicx}
\usepackage{epsfig}
\usepackage{color}
\usepackage{multirow}

\newcommand{\btopkg}{B \rightarrow \phi K \gamma}
\newcommand{\btopkog}{B^0 \rightarrow \phi K^0 \gamma}
\newcommand{\btopksg}{B^0 \rightarrow \phi K_S^0 \gamma}
\newcommand{\btopkpg}{B^+ \rightarrow \phi K^+ \gamma}

\begin{document}

  \title{First Observation of Radiative {\boldmath $B^0 \rightarrow \phi K^0 \gamma$} Decays and Measurements of Their Time-Dependent {\boldmath $CP$} Violation}  

%%%%%%%% Start of author list

%%% Paper:    B -> phi K gamma
%%% Journal:  Physical Review Letters
%%% Contacts: H. Sahoo (himansu@phys.hawaii.edu)
%%%           T. E. Browder (teb@phys.hawaii.edu)
%%% Non-responding authors or those who said NO are commented out.
%%% ====================================================================
%%% Click the RELOAD button on your web browser to see the updated file.
%%% ====================================================================
%%% Use \input{author} to insert this material into your latex file.
%%%%% Force institutions to appear in alphabetical order when typeset.
%%%\affiliation{University of Bonn, Bonn}
\affiliation{Budker Institute of Nuclear Physics, Novosibirsk}
\affiliation{Faculty of Mathematics and Physics, Charles University, Prague}
%%%\affiliation{Chiba University, Chiba}
\affiliation{University of Cincinnati, Cincinnati, Ohio 45221}
%%%\affiliation{Department of Physics, Fu Jen Catholic University, Taipei}
\affiliation{Justus-Liebig-Universit\"at Gie\ss{}en, Gie\ss{}en}
\affiliation{Gifu University, Gifu}
%%%\affiliation{The Graduate University for Advanced Studies, Hayama}
%%%\affiliation{Gyeongsang National University, Chinju}
\affiliation{Hanyang University, Seoul}
\affiliation{University of Hawaii, Honolulu, Hawaii 96822}
\affiliation{High Energy Accelerator Research Organization (KEK), Tsukuba}
%%%\affiliation{Hiroshima Institute of Technology, Hiroshima}
%%%\affiliation{University of Illinois at Urbana-Champaign, Urbana, Illinois 61801}
\affiliation{Indian Institute of Technology Guwahati, Guwahati}
\affiliation{Indian Institute of Technology Madras, Madras}
%%%\affiliation{Indiana University, Bloomington, Indiana 47408}
\affiliation{Institute of High Energy Physics, Chinese Academy of Sciences, Beijing}
\affiliation{Institute of High Energy Physics, Vienna}
\affiliation{Institute of High Energy Physics, Protvino}
%%%\affiliation{Institute of Mathematical Sciences, Chennai}
%%%\affiliation{INFN - Sezione di Torino, Torino}
\affiliation{Institute for Theoretical and Experimental Physics, Moscow}
\affiliation{J. Stefan Institute, Ljubljana}
\affiliation{Kanagawa University, Yokohama}
\affiliation{Institut f\"ur Experimentelle Kernphysik, Karlsruher Institut f\"ur Technologie, Karlsruhe}
\affiliation{Korea Institute of Science and Technology Information, Daejeon}
\affiliation{Korea University, Seoul}
%%%\affiliation{Kyoto University, Kyoto}
\affiliation{Kyungpook National University, Taegu}
\affiliation{\'Ecole Polytechnique F\'ed\'erale de Lausanne (EPFL), Lausanne}
\affiliation{Faculty of Mathematics and Physics, University of Ljubljana, Ljubljana}
\affiliation{University of Maribor, Maribor}
\affiliation{Max-Planck-Institut f\"ur Physik, M\"unchen}
\affiliation{University of Melbourne, School of Physics, Victoria 3010}
\affiliation{Nagoya University, Nagoya}
%%%\affiliation{Nara University of Education, Nara}
\affiliation{Nara Women's University, Nara}
\affiliation{National Central University, Chung-li}
%%%\affiliation{National United University, Miao Li}
\affiliation{Department of Physics, National Taiwan University, Taipei}
\affiliation{H. Niewodniczanski Institute of Nuclear Physics, Krakow}
\affiliation{Nippon Dental University, Niigata}
\affiliation{Niigata University, Niigata}
\affiliation{University of Nova Gorica, Nova Gorica}
\affiliation{Novosibirsk State University, Novosibirsk}
\affiliation{Osaka City University, Osaka}
%%%\affiliation{Osaka University, Osaka}
\affiliation{Pacific Northwest National Laboratory, Richland, Washington 99352}
\affiliation{Panjab University, Chandigarh}
%%%\affiliation{Peking University, Beijing}
%%%\affiliation{Princeton University, Princeton, New Jersey 08544}
\affiliation{Research Center for Nuclear Physics, Osaka}
%%%\affiliation{RIKEN BNL Research Center, Upton, New York 11973}
\affiliation{Saga University, Saga}
\affiliation{University of Science and Technology of China, Hefei}
\affiliation{Seoul National University, Seoul}
%%%\affiliation{Shinshu University, Nagano}
\affiliation{Sungkyunkwan University, Suwon}
\affiliation{School of Physics, University of Sydney, NSW 2006}
\affiliation{Tata Institute of Fundamental Research, Mumbai}
\affiliation{Excellence Cluster Universe, Technische Universit\"at M\"unchen, Garching}
%%%\affiliation{Toho University, Funabashi}
\affiliation{Tohoku Gakuin University, Tagajo}
\affiliation{Tohoku University, Sendai}
\affiliation{Department of Physics, University of Tokyo, Tokyo}
\affiliation{Tokyo Institute of Technology, Tokyo}
\affiliation{Tokyo Metropolitan University, Tokyo}
\affiliation{Tokyo University of Agriculture and Technology, Tokyo}
%%%\affiliation{Toyama National College of Maritime Technology, Toyama}
\affiliation{CNP, Virginia Polytechnic Institute and State University, Blacksburg, Virginia 24061}
\affiliation{Wayne State University, Detroit, Michigan 48202}
%%%\affiliation{Yamagata University, Yamagata}
\affiliation{Yonsei University, Seoul}
  \author{H.~Sahoo}\affiliation{University of Hawaii, Honolulu, Hawaii 96822} % Hawaii
  \author{T.~E.~Browder}\affiliation{University of Hawaii, Honolulu, Hawaii 96822} % Hawaii
  \author{I.~Adachi}\affiliation{High Energy Accelerator Research Organization (KEK), Tsukuba} % KEK
% \author{K.~Adamczyk}\affiliation{H. Niewodniczanski Institute of Nuclear Physics, Krakow} % Krakow
% \author{H.~Aihara}\affiliation{Department of Physics, University of Tokyo, Tokyo} % Tokyo
% \author{K.~Arinstein}\affiliation{Budker Institute of Nuclear Physics, Novosibirsk}\affiliation{Novosibirsk State University, Novosibirsk} % BINP
% \author{Y.~Arita}\affiliation{Nagoya University, Nagoya} % Nagoya
  \author{D.~M.~Asner}\affiliation{Pacific Northwest National Laboratory, Richland, Washington 99352} % PNNL
% \author{T.~Aso}\affiliation{Toyama National College of Maritime Technology, Toyama} % Toyama
  \author{V.~Aulchenko}\affiliation{Budker Institute of Nuclear Physics, Novosibirsk}\affiliation{Novosibirsk State University, Novosibirsk} % BINP
% \author{T.~Aushev}\affiliation{Institute for Theoretical and Experimental Physics, Moscow} % ITEP
% \author{T.~Aziz}\affiliation{Tata Institute of Fundamental Research, Mumbai} % Tata
  \author{A.~M.~Bakich}\affiliation{School of Physics, University of Sydney, NSW 2006} % Sydney
% \author{Y.~Ban}\affiliation{Peking University, Beijing} % Peking
  \author{E.~Barberio}\affiliation{University of Melbourne, School of Physics, Victoria 3010} % Melbourne
% \author{A.~Bay}\affiliation{\'Ecole Polytechnique F\'ed\'erale de Lausanne (EPFL), Lausanne} % Lausanne
% \author{I.~Bedny}\affiliation{Budker Institute of Nuclear Physics, Novosibirsk}\affiliation{Novosibirsk State University, Novosibirsk} % BINP
% \author{M.~Belhorn}\affiliation{University of Cincinnati, Cincinnati, Ohio 45221} % Cincinnati
  \author{K.~Belous}\affiliation{Institute of High Energy Physics, Protvino} % Protvino
  \author{V.~Bhardwaj}\affiliation{Panjab University, Chandigarh} % Panjab
  \author{B.~Bhuyan}\affiliation{Indian Institute of Technology Guwahati, Guwahati} % IITG
% \author{M.~Bischofberger}\affiliation{Nara Women's University, Nara} % Nara
% \author{S.~Blyth}\affiliation{National United University, Miao Li} % NUU
% ...
% Karlsruhe
% \author{N.~Zwahlen}\affiliation{\'Ecole Polytechnique F\'ed\'erale de Lausanne (EPFL), Lausanne} % Lausanne
% \author{O.~Zyukova}\affiliation{Budker Institute of Nuclear Physics, Novosibirsk}\affiliation{Novosibirsk State University, Novosibirsk} % BINP
\collaboration{The Belle Collaboration}

%%%%%%% End of author list

\noaffiliation
  
\begin{abstract}
We report the first observation of the radiative decay $\btopkog$ 
using a data sample 
of $772 \times 10^6$ $B\overline{B}$ pairs collected at 
the $\Upsilon(4S)$ resonance 
with the Belle detector at the KEKB asymmetric-energy $e^+e^-$ collider. 
We observe a signal of 
$37\pm8$ 
events with a significance of 
$5.4$ standard deviations
including systematic uncertainties. The measured branching fraction is 
${\cal B}(\btopkog) = (2.74\pm 0.60 \pm 0.32) \times 10^{-6}$, where
the uncertainties are statistical and systematic, respectively.
We also report the first measurements of time-dependent 
$CP$ violation parameters:
${\mathcal S}_{\phi K_S^0 \gamma} = +0.74^{+0.72}_{-1.05} (\rm{stat})^{+0.10}_{-0.24} (\rm{syst})$ and
${\mathcal A}_{\phi K_S^0 \gamma} = +0.35 \pm 0.58 (\rm{stat})^{+0.23}_{-0.10} (\rm{syst})$.
Furthermore, we measure
${\mathcal B}(\btopkpg) = (2.48\pm 0.30 \pm 0.24) \times 10^{-6}$,
${\mathcal A}_{CP} = -0.03\pm 0.11\pm 0.08$
and find that the signal is concentrated in the
$M_{\phi K}$ mass region near threshold.
\end{abstract}


\maketitle

%%%%%%%%%%%%%%%%%%%%%%%%%
%% Motivation
%%%%%%%%%%%%%%%%%%%%%%%%%
\par Rare radiative $B$ meson decays play an important role in the 
search for physics 
beyond the standard model (SM).
These are flavor changing neutral current decays,
forbidden at tree level in the SM but allowed through
electroweak loop processes. 
The loop can be mediated by non-SM particles
(for example, charged Higgs or SUSY particles)
and therefore is sensitive to new physics (NP).
Here we report
the first observation of a new
$b\rightarrow s$ radiative penguin decay mode, $\btopkog$,
as well as measurements of its time-dependent $CP$ asymmetry.
This type of decay is sensitive to NP
from right-handed currents~\cite{ags} 
and will be useful for precise time-dependent 
measurements at future
high-luminosity flavor facilities~\cite{SuperKEKB,SuperB,LHCb}.
%%%%%%%%%%%%%%%%%%%%%%%%%
% Branching fraction Motivation
%%%%%%%%%%%%%%%%%%%%%%%%%
\par The emitted photons 
in $b \rightarrow s \gamma$ 
($\overline{b} \rightarrow \overline{s} \gamma$) decays
are predominantly left-handed (right-handed) in the SM, and
hence the time-dependent $CP$ asymmetry 
is suppressed by the quark mass ratio ($2m_s/m_b$).
The expected mixing-induced $CP$ asymmetry parameter (${\cal S}$)
is ${\cal O}(3\%)$
and the direct $CP$ asymmetry parameter 
(${\cal A}$) is $\sim 0.6\%$~\cite{ags}.
In several extensions of the SM, both photon helicities can
contribute to the decay.
Therefore, any significantly larger $CP$ asymmetry 
would be clear evidence for NP.
In contrast to 
$B^0 \to K^{*0} (\to K_S^0 \pi^0) \gamma$~\cite{belle-kstg,babar-kstg}, 
another related mode that is sensitive to NP, 
the time dependence of 
$\btopksg$ can be measured from the $\phi \to K^+K^-$ decay
and does not require a difficult measurement of the 
long lived $K_S^0$ decay inside the 
%silicon 
inner tracking volume or 
reconstruction of a low energy $\pi^0$.
The $\btopkg$
mode can be used to search for 
a possible contribution from 
kaonic resonances decaying to $\phi K$. 
Furthermore, we can also probe the photon polarization 
using the angular distributions of 
the final state hadrons~\cite{pol}.
%%%%%%%%%%%%%%%%%%%%%%%%%
%% History
%%%%%%%%%%%%%%%%%%%%%%%%%
\par 
Results on $\btopkg$ decays have been reported 
by both Belle and BaBar collaborations based on
$96 \times 10^6$ $B\overline{B}$~\cite{alex_prl} and 
$228 \times 10^6$ $B\overline{B}$~\cite{phikgamma_babar} pairs,
respectively.
Only upper limits on
$\mathcal{B}(\btopkog)$ were given.
Here we report the first observation of $\btopkog$, 
the first measurements of time-dependent $CP$ violation 
in this mode, as well as 
more precise measurements 
of $\btopkpg$~\cite{conj}.
The data set used consists of
$772 \times 10^6$ $B\overline{B}$ pairs
collected at the $\Upsilon(4S)$ resonance 
with the Belle detector~\cite{Belle} 
at the KEKB asymmetric-energy $e^+e^-$ (3.5 on 8.0 GeV) 
collider~\cite{kekb}.
%%%%%%%%%%%%%%%%%%%%%%%%%
%% TCPV PDF
%%%%%%%%%%%%%%%%%%%%%%%%%
\par At KEKB, the $\Upsilon(4S)$ is produced with a Lorentz 
boost of $\beta \gamma = 0.425$ along the $z$ axis, which 
is defined as opposite to the $e^+$ beam direction.
In the decay chain 
$\Upsilon(4S) \rightarrow B^0 \overline{B}{}^0 \rightarrow f_{\rm rec}f_{\rm tag}$, 
where one of the $B$ mesons decays at 
time $t_{\rm rec}$ to the signal mode $f_{\rm rec}$ and the other decays
at time $t_{\rm tag}$ to a final state $f_{\rm tag}$ that distinguishes
between $B^0$ and $\overline{B}{}^0$, the decay rate has a time 
dependence given by
\begin{eqnarray}
{\mathcal P}(\Delta{t}) \:=\: \frac{ e^{-|\Delta{t}|/{\tau_{B^0}}} }{4\tau_{B^0}}
\biggl\{1 &+& q \,
 \Bigl[ {\mathcal S} \sin(\Delta m_d \Delta{t})  \nonumber \\
 &+& {\mathcal A} \cos(\Delta m_d \Delta{t})
\Bigr] \biggr\}.
\label{eq_decay}
\end{eqnarray}
Here $\tau_{B^0}$ is the neutral $B$ lifetime,
$\Delta m_d$ is the mass difference between the two 
neutral $B$ mass eigenstates, 
$\Delta t = t_{\rm rec} - t_{\rm tag}$, and the $b$-flavor charge 
$q$ equals $+1$ $(-1)$ when the tagging $B$ meson is a 
$B^0$ ($\overline{B}{}^0$). 
Since the $B^0$ and $\overline{B}{}^0$ are approximately 
at rest in the $\Upsilon(4S)$ center-of-mass system (cms),
$\Delta t$ can be determined from $\Delta z$, 
the displacement in $z$ between the two 
decay vertices: $\Delta t \simeq \Delta z / (\beta \gamma c)$.
%%%%%%%%%%%%%%%%%%%%%%%%%
%% Reconstruction
%%%%%%%%%%%%%%%%%%%%%%%%%
\par Signal candidates are reconstructed in the 
$\btopkpg$ and $\btopksg$ modes, with 
$\phi \rightarrow K^+K^-$ and $K_S^0 \rightarrow \pi^+\pi^-$.
Charged kaons are identified by requiring a likelihood ratio
$\mathcal{L}_{K/\pi} 
\left[=\mathcal{L}_{K}/(\mathcal{L}_{K}+\mathcal{L}_{\pi})\right]> 0.6$,
which is calculated using information from the 
aerogel Cherenkov, time-of-flight, and drift chamber
detectors.
This requirement has an efficiency of $90\%$ for kaons 
and an $8\%$ pion fake rate.
A less restrictive selection
$\mathcal{L}_{K/\pi} > 0.4$ 
is applied to the kaon candidates that are used to 
reconstruct the $\phi$ meson.
The invariant mass of the $\phi$ candidates is required to satisfy
$\left\vert M_{K^+K^-}-m_{\phi}\right\vert < 10$ MeV/$c^2$, 
where $m_{\phi}$ denotes the $\phi$ meson world-average mass~\cite{pdg}.
The $K_S^0$ selection criteria are the same as those 
described in Ref.~\cite{belle_b2s}; the invariant mass of the pion
pairs should be in the range
$M_{\pi^+\pi^-} \in \left[482, 514\right]$ MeV/$c^2$.
The high energy prompt photons must lie in
the barrel region of the
calorimeter (ECL), have a cms energy
$E_{\gamma}^{\rm cms} \in \left[1.4,3.4\right]$ GeV and
a shower shape consistent with that of a photon.
We also suppress the background photons
from $\pi^0$($\eta$) $\rightarrow \gamma \gamma$ 
using a likelihood $\mathcal{L}_{\pi^0}$($\mathcal{L}_{\eta}$) $< 0.25$, 
as described in Ref.~\cite{pi0etaveto}.
%%%%%%%%%%%%%%%%%%%%%%%%%
%% B meson reconstruction
%%%%%%%%%%%%%%%%%%%%%%%%%
\par We combine a $\phi$ meson candidate, a charged or neutral 
kaon candidate, and a radiative photon to form a $B$ meson. 
$B$ candidates are identified using 
two kinematic variables: the energy difference 
$\Delta E \equiv E_B^{\rm cms} - E_{\rm beam}^{\rm cms}$ and the
beam-energy-constrained mass 
$M_{\rm bc} \equiv \sqrt{(E_{\rm beam}^{\rm cms}/c^2)^2 - (p_B^{\rm cms}/c)^2}$,
where $E_{\rm beam}^{\rm cms}$ is the beam energy in the cms, and 
$E_B^{\rm cms}$ and $p_B^{\rm cms}$ are the cms energy and momentum, 
respectively, of the reconstructed $B$ candidate.
In the $M_{\rm bc}$ calculation, the photon momentum is replaced by
$(E_{\rm beam}^{\rm cms} - E_{\phi K}^{\rm cms})$ 
to improve its resolution.
The candidates
that satisfy the requirements
$M_{\rm bc} > 5.2 \;{\rm GeV/}c^2$ and 
$\left|\Delta E\right| < 0.3 \;\rm{GeV}$ 
are retained for further analysis.
Using Monte Carlo (MC) simulations, we find 
nearly $12\%$ ($3\%$) of signal events 
in the charged (neutral) mode have more than one $B$ candidate.
In case of multiple candidates, we
choose the best candidate based on a series of 
selection criteria, 
which depend on a $\chi^2$ variable formed using 
the candidate's $\phi$ mass
(and the $K_S^0$ mass in the neutral mode)
as well as
the highest $E_{\gamma}^{\rm cms}$ and  
the highest $\mathcal{L}_{K/\pi}$ in the charged mode.
For events with multiple candidates, this selection method chooses the
correct $B$ candidate for the 
charged (neutral) mode
$57\%$ ($69\%$) of the time.
%%%%%%%%%%%%%%%%%%%%%%%%%
%% Continuum background suppression
%%%%%%%%%%%%%%%%%%%%%%%%%
\par The dominant background comes from 
$e^+e^- \rightarrow q\overline{q}$
($q = u, d, s, c$) 
continuum events. 
We use two event-shape variables 
(a Fisher discriminant formed from
modified Fox-Wolfram moments~\cite{fisher} and
the cosine of the angle between the
$B$ flight direction and the beam axis, $\cos \theta_B$, 
in the cms frame) 
to distinguish spherically 
symmetric $B\overline{B}$ events from the jet-like continuum background.
From these variables we form a likelihood ratio, denoted by
$\mathcal{R}_{s/b}$.
We require
$\mathcal{R}_{s/b}>0.65$, which removes $91\%$ 
of the continuum while retaining $76\%$ of the signal.
%%%%%%%%%%%%%%%%%%%%%%%%%
%% Generic and Rare background suppression
%%%%%%%%%%%%%%%%%%%%%%%%%
In addition to the continuum,
various $B\overline{B}$
background sources are also studied. 
In the $\btopksg$ mode,
backgrounds from some $b \rightarrow c$ decays such as
$D^0\pi^0$, $D^0\eta$ and $D^-\rho^+$,
peak in the $M_{\rm bc}$ distribution. 
We remove the dominant peaking backgrounds by applying a veto 
to $\phi K_S^0$ combinations consistent 
within detector resolution ($\pm 4\,\sigma$)
with the nominal $D$ mass~\cite{pdg}.
Some of the charmless backgrounds, where the $B$ meson decays to
$\phi K^{*}(892)$, $\phi K \pi^0$ and $\phi K \eta$ 
also peak in $M_{\rm bc}$ but shift towards lower $\Delta E$. 
Another significant background is
non-resonant (NR) $B \rightarrow K^+ K^- K \gamma$, which peaks in the
$\Delta E$-$M_{\rm bc}$ signal region; it is estimated using
the $\phi$ mass sideband,
$M_{K^+K^-} \in \left[1.05,1.30\right]$ GeV/$c^2$, in data.
%%%%%%%%%%%%%%%%%%%%%%%%%%%%%
%% DE-Mbc fit distribution
%%%%%%%%%%%%%%%%%%%%%%%%%%%%%
\begin{figure}[htbp]
  \begin{center}
    \resizebox{0.51\columnwidth}{!}{\includegraphics{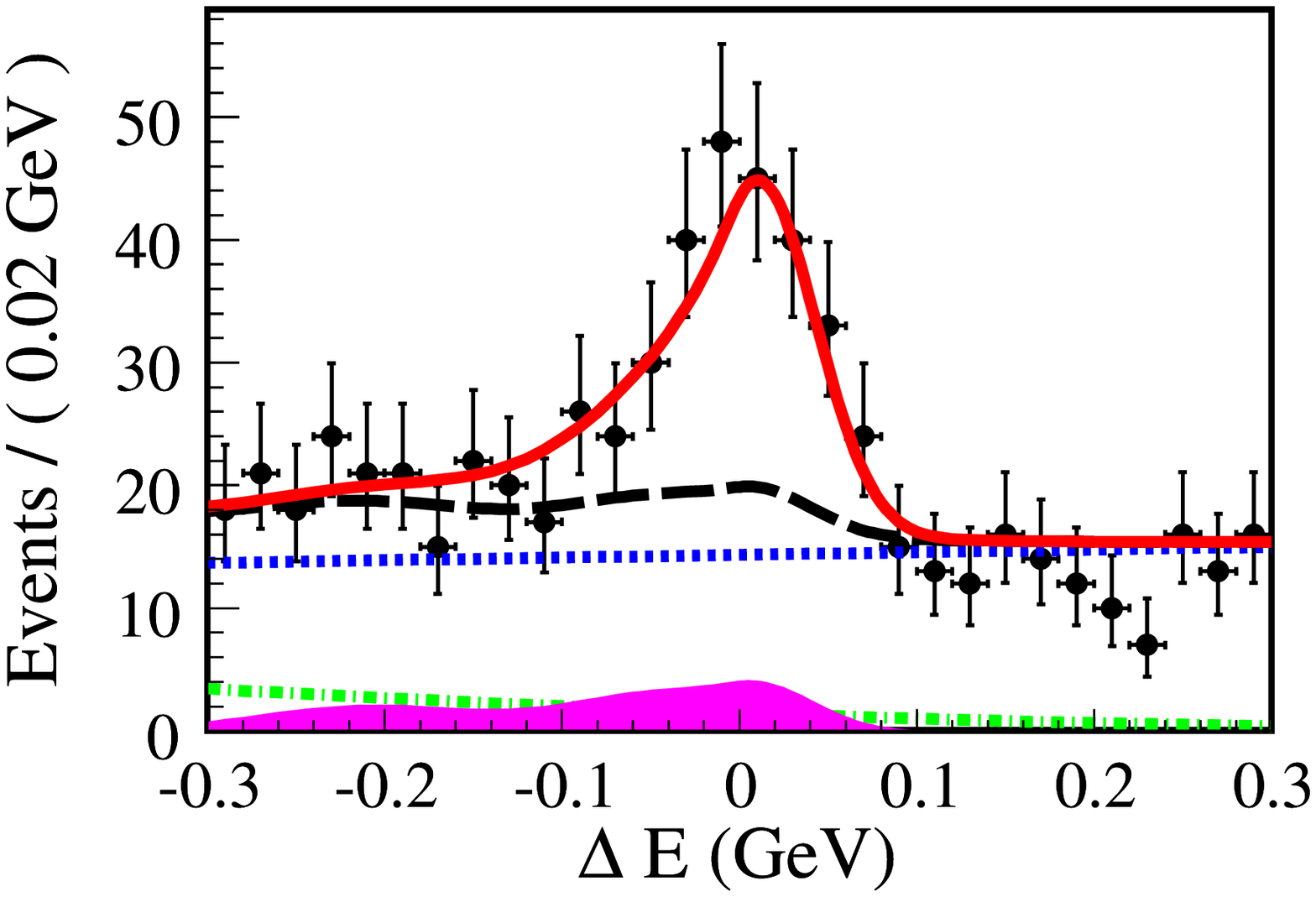}}
    \hspace{-4mm}\resizebox{0.51\columnwidth}{!}{\includegraphics{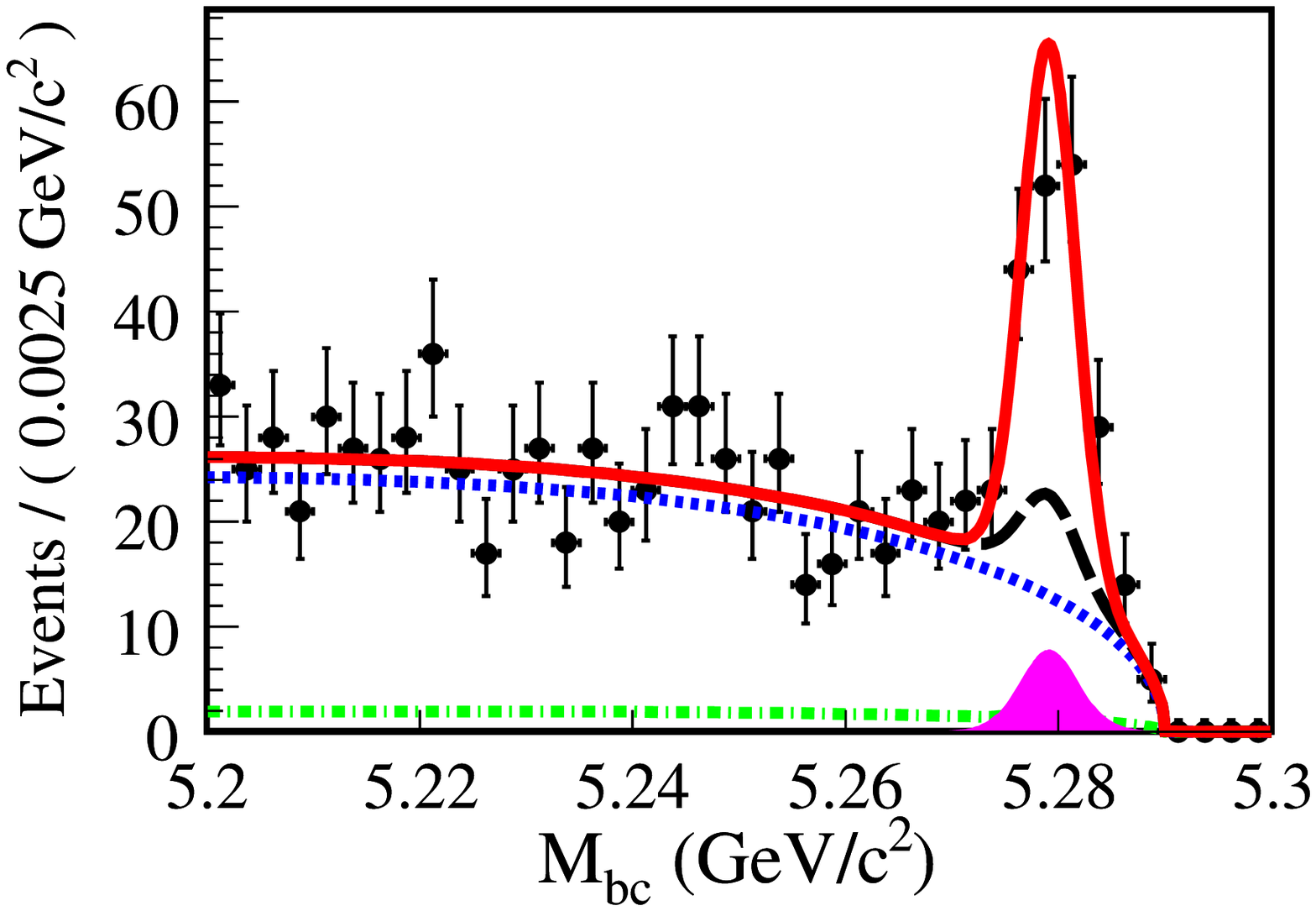}}
    \resizebox{0.51\columnwidth}{!}{\includegraphics{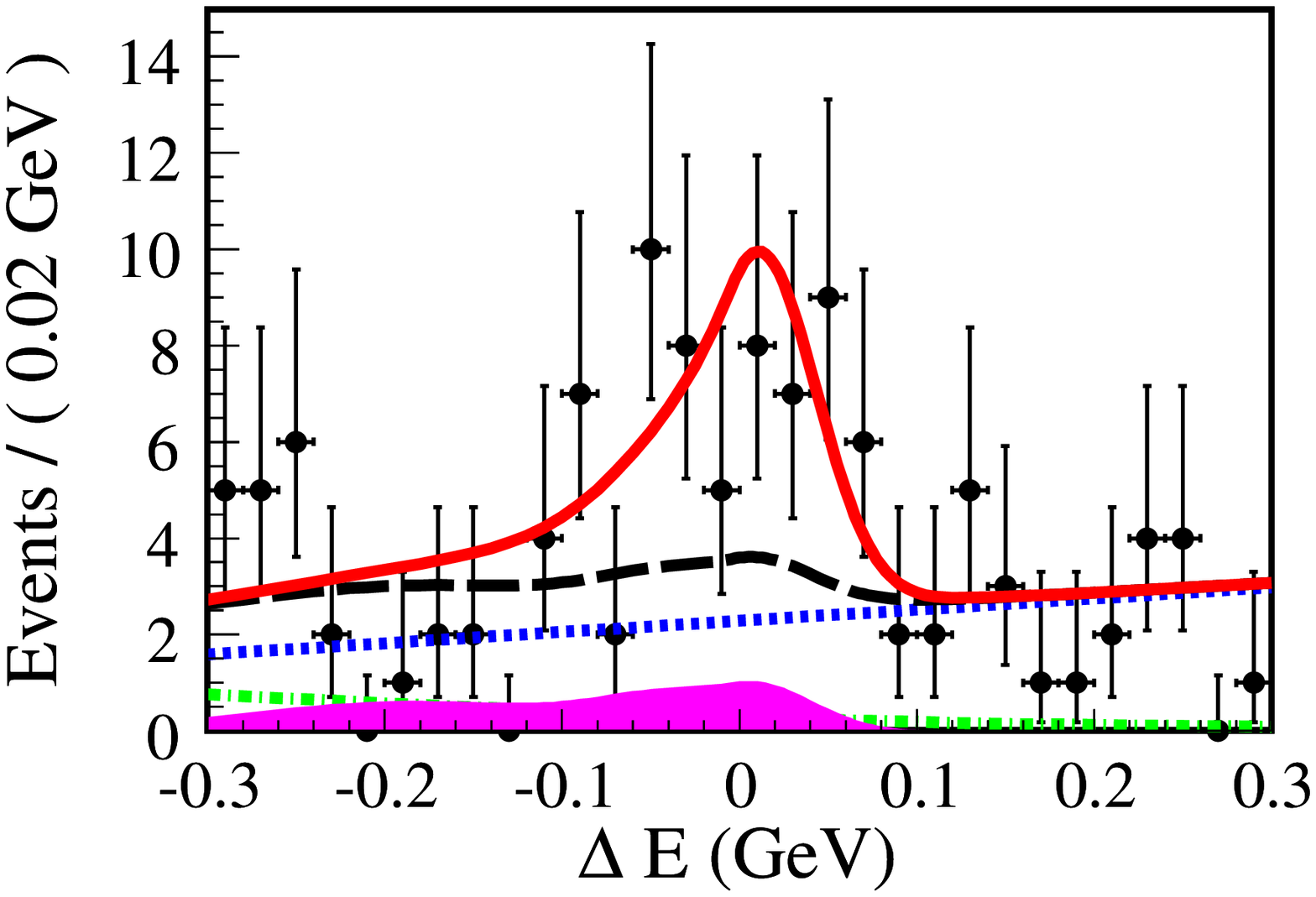}}
    \hspace{-4mm}\resizebox{0.51\columnwidth}{!}{\includegraphics{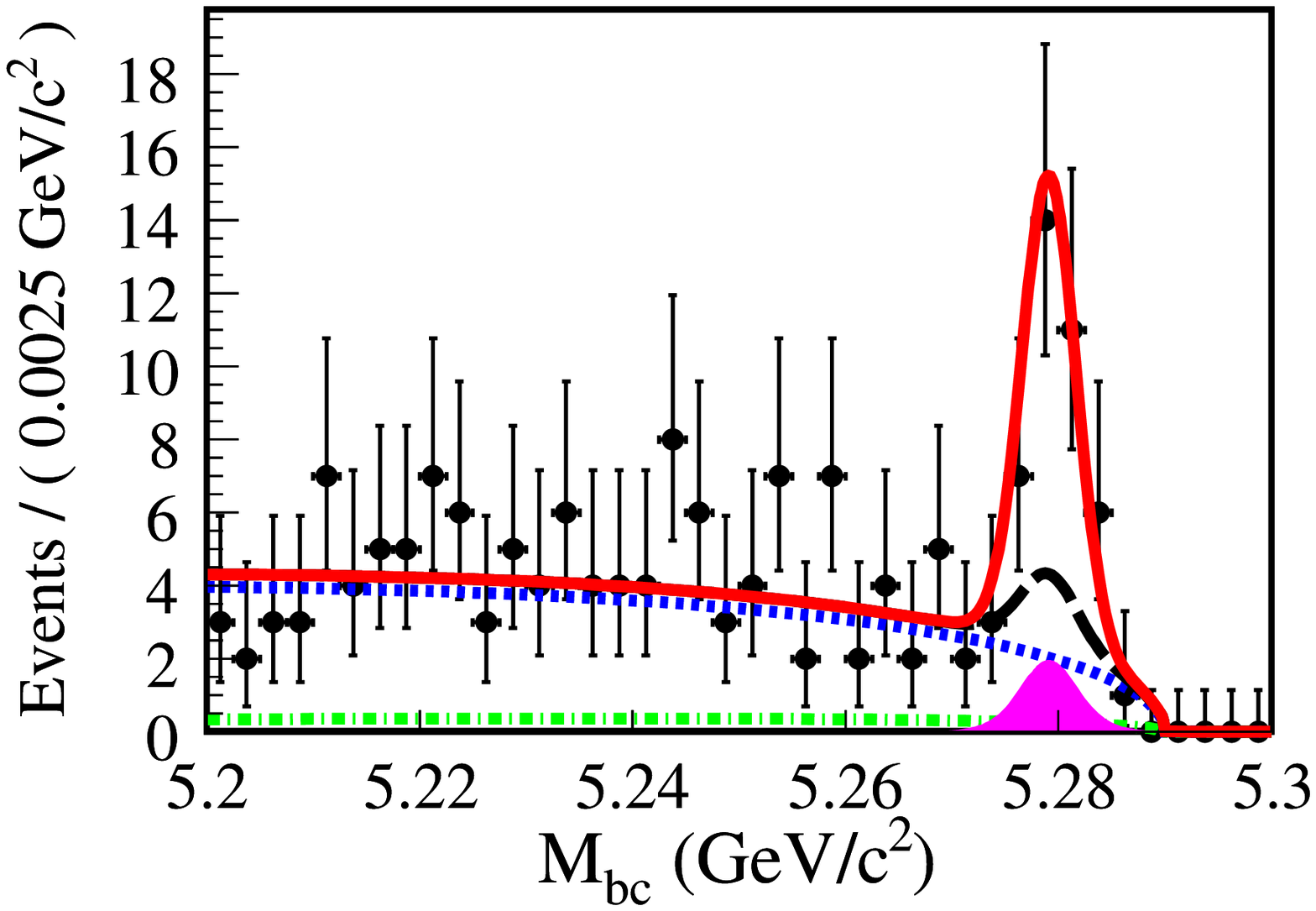}}
    \caption{$\Delta E$ and $M_{\rm bc}$ projections 
      for $\btopkpg$ (upper) and $\btopksg$ (lower). 
      The $\Delta E$ projections include the requirement 
      $M_{\rm bc} \in \left[5.27,5.29\right]$ GeV/$c^2$ while the 
      $M_{\rm bc}$ projections require $\Delta E \in \left[-0.08,0.05\right]$ GeV.
      The points with error bars are the data. The curves show the total fit function (solid red), 
      total background function (long-dashed black), continuum component (dotted blue), 
      the $b\rightarrow c$ component (dashed-dotted green) and the non-resonant 
      component as well as other charmless backgrounds (filled magenta histogram).
    }
    \label{fig:dembc}
  \end{center}
\end{figure}
\par The signal yield is obtained from an extended
unbinned maximum-likelihood (UML) fit to the two-dimensional 
$\Delta E$-$M_{\rm bc}$ distribution.
We model the shape for the signal component using the product of a 
Crystal Ball line shape~\cite{cbshape} for $\Delta E$ and 
a Gaussian for $M_{\rm bc}$. 
The continuum background is represented by the product of a
first-order polynomial
for $\Delta E$ and an ARGUS~\cite{argus} function for $M_{\rm bc}$.
The $b \rightarrow c$ background is described by the product of a
second-order polynomial
for $\Delta E$ and the sum of an
ARGUS and a Gaussian function for $M_{\rm bc}$. 
For the small charmless backgrounds 
(except for the NR component), we use
the sum of two Gaussians for $\Delta E$ and a Gaussian for $M_{\rm bc}$.
The probability density function (PDF)
is the product of these 
two functional forms~\cite{charmless-pdf}.
In the final fit the continuum parameters are allowed to vary 
while all other
background parameters are fixed to the values from MC simulations. 
The shapes of the $b \to c$ and NR
peaking background components are fixed to that of the signal.
In the charged mode, the NR 
background yield, $(12.5\pm6.7)\%$ of the signal,
is fixed from the $\phi$ mass sideband. 
Since the neutral mode is limited by statistics,
we assume isospin symmetry and use the same NR fraction.
The signal shapes are  adjusted for small differences 
between MC simulations and data using a
$B^0 \rightarrow K^{*}(892)^0(\rightarrow K^+\pi^-) \gamma$ 
control sample, with
$M_{K^+\pi^-} \in \left[820,970\right]$ MeV/$c^2$. 
The fit yields a signal of 
$144\pm17$ $\btopkpg$ and $37\pm8$ $\btopksg$ events.
The projections of the fit results onto $\Delta E$ 
and $M_{\rm bc}$ are shown in Fig.~\ref{fig:dembc}.
The signal significance is defined as 
$\sqrt{-2\,\ln({\cal L}_0/{\cal L}_{\rm max})}$,
where ${\cal L}_{\rm max}$ is
the maximum likelihood for the nominal fit and ${\cal L}_0$ 
is the corresponding value with the signal yield fixed to zero.
The additive sources of systematic uncertainty 
(described below) are
included in the significance by 
varying each by its error and taking the lowest significance.
The signal in the charged mode has a significance of
$9.6\,\sigma$, whereas that for the neutral mode is $5.4\,\sigma$.
%%%%%%%%%%%%%%%%%%%%%%%%%
%% phiK spectra
%%%%%%%%%%%%%%%%%%%%%%%%%
\begin{figure}[htbp]
  \begin{center}
    \resizebox{0.52\columnwidth}{!}{\includegraphics{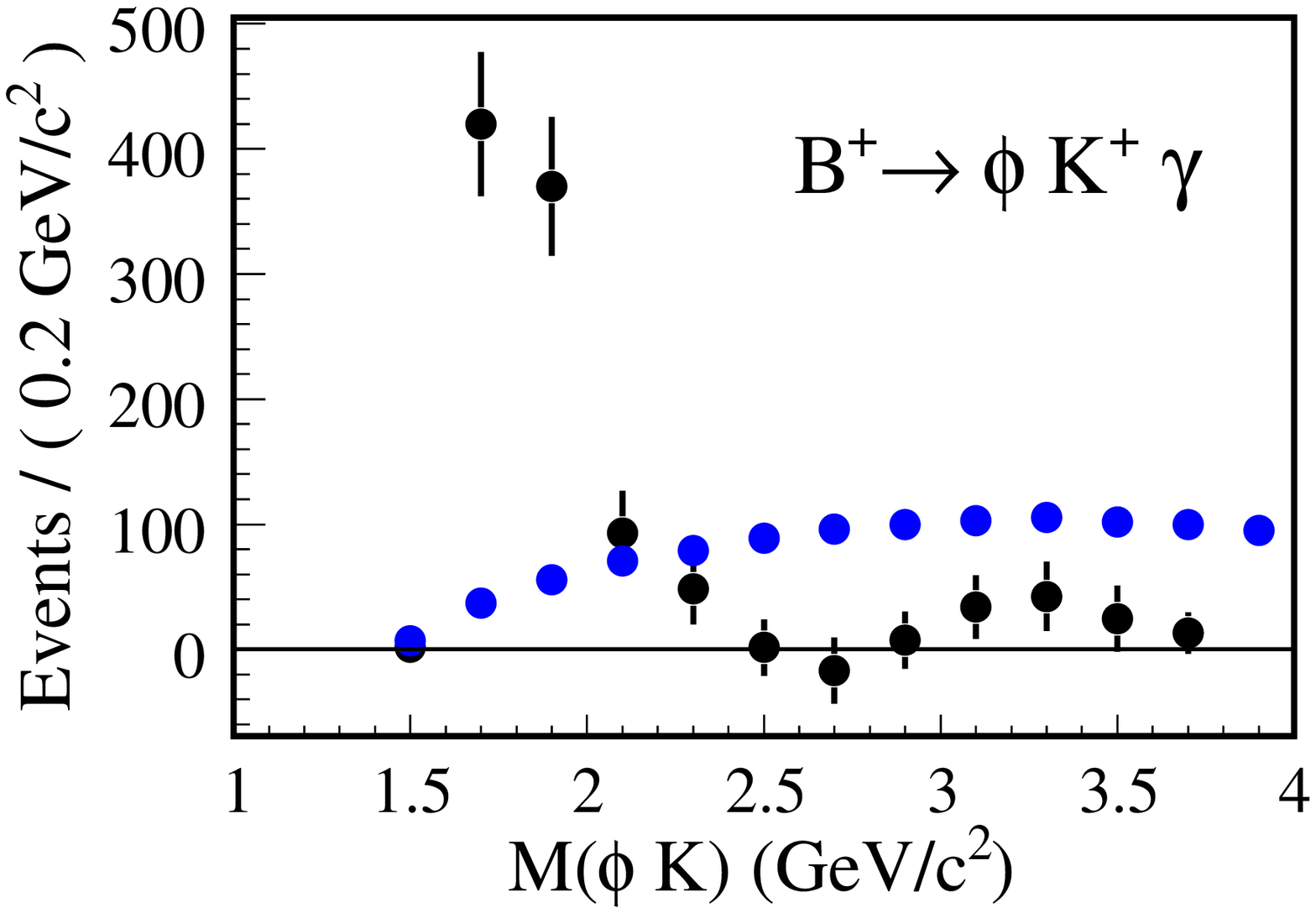}}
    \hspace{-5mm}\resizebox{0.52\columnwidth}{!}{\includegraphics{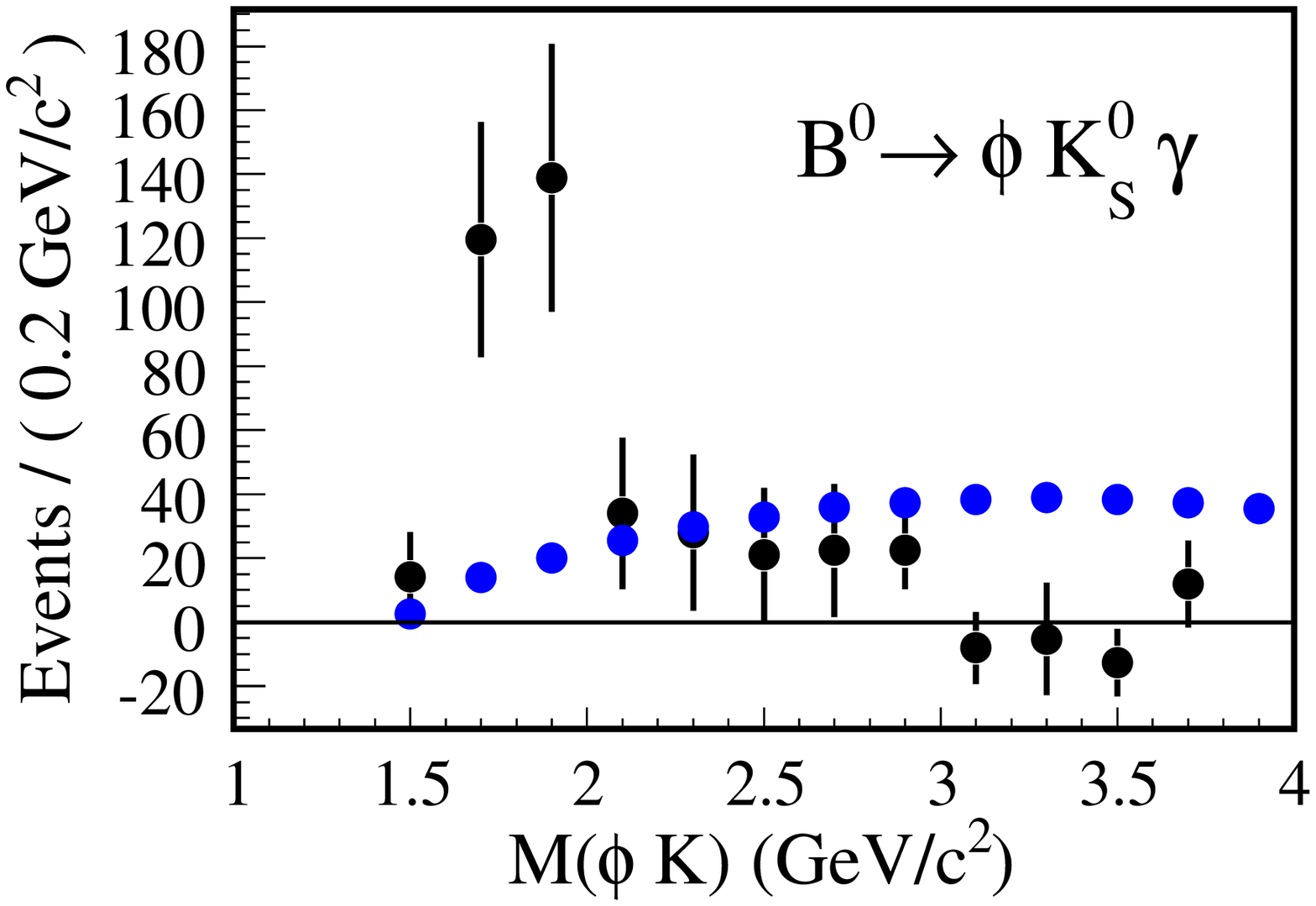}}
    \caption{Background-subtracted and efficiency-corrected 
      $\phi K$ mass distributions for 
      the charged (left) and neutral (right) modes. 
      The points with error bars 
      represent the data. The yield in each bin is obtained by 
      the fitting procedure described in the text. 
      A three-body phase-space model from MC simulation 
      is shown by the filled circles (blue points) and 
      normalized to the total 
      data signal yield.}
    \label{fig:mxs}
  \end{center}
\end{figure}

%%%%%%%%%%%%%%%%%%%%%%%%%%%%%
%% M(Xs) mass distribution
%%%%%%%%%%%%%%%%%%%%%%%%%%%%%
\par To measure the $M_{\phi K}$ distribution,
we repeat the fit in bins of $\phi K$ mass and the resulting signal yields
are corrected for the detection efficiency.
Nearly $72\%$ of the signal events are concentrated
in the low-mass region,
$M_{\phi K} \in \left[1.5,2.0\right]$ GeV/$c^2$, 
as shown in Fig.~\ref{fig:mxs}.
The MC efficiencies 
are reweighted according to this $M_{\phi K}$ dependence.
These spectra are consistent with the expectations from
the pQCD model for non-resonant $\btopkg$ decays~\cite{hsiangnan}.
With the present statistics no clear evidence is found for the existence
of a kaonic resonance decaying to $\phi K$.
 
\par From the signal yield ($N_{\rm sig}$), we calculate the 
branching fraction ($\mathcal{B}$) as 
$N_{\rm sig}/(\epsilon \times N_{B\overline{B}} \times \mathcal{B}_{\rm sec})$, 
where $\epsilon$ is the weighted efficiency
[$(15.3\pm0.1 (\mathrm{stat}))\%$ for the charged mode and 
$(10.0\pm0.1 (\mathrm{stat}))\%$ for the neutral mode], 
$N_{B\overline{B}}$ is the number of $B\overline{B}$ 
pairs in the data sample, and $\mathcal{B}_{\rm sec}$ 
is the product of daughter branching fractions~\cite{pdg}.
We obtain
${\mathcal B}(\btopkpg) = (2.48\pm 0.30 \pm 0.24) \times 10^{-6}$ and
${\mathcal B}(\btopkog) = (2.74\pm 0.60 \pm 0.32) \times 10^{-6}$,
where the uncertainties are statistical and systematic, respectively.

%%%%%%%%%%%%%%%%%%%%%%%%%%%%%%
%% Systematic Errors for BF
%%%%%%%%%%%%%%%%%%%%%%%%%%%%%%
\par We evaluate the systematic uncertainties on the signal yield
by fitting the data with each fixed parameter
varied by its $\pm 1\,\sigma$ error, 
and then taking the quadratic sum of all differences from the nominal 
value.
The largest contribution of $8.0\%$ arises from the NR yield.
The other sources of systematic error are from
charged track efficiency ($\sim 1.1\%$ per track),
photon detection efficiency ($2.4\%$), 
particle identification ($1.4\%$), 
number of produced $B\overline{B}$ pairs ($1.4\%$),
$\phi$ and $K_S^0$ branching fractions ($1.2\%$),
$K_S^0$ reconstruction ($4.6\%$),
and the requirement on $\mathcal{R}_{s/b}$ ($0.3\%$).
The statistical uncertainty on the MC efficiency 
after reweighting is 
$1.0\%$ ($1.2\%$) in the charged (neutral) mode.
Furthermore, we assign a systematic error of $0.2\%$ ($2.7\%$)
for possible fit bias, which is obtained from
ensemble tests with MC pseudo-experiments.
The total systematic uncertainty on the
branching fraction is
$9.5\%$ ($11.7\%$).
%%%%%%%%%%%%%%%%%%%%%%%%%%%%%%
%% CP violation measurement
%%%%%%%%%%%%%%%%%%%%%%%%%%%%%%
\par For the $CP$ asymmetry fit, we select events in the
signal region defined as
$M_{\rm bc} \in \left[5.27,5.29\right]$ GeV/$c^2$ and 
$\Delta E \in \left[-0.2,0.1\right]$ GeV.
Different selection criteria on 
$\mathcal{R}_{s/b}$ are used depending upon the 
flavor-tagging information.
In addition, ECL endcap region photons are included in the analysis.
We use a flavor tagging algorithm~\cite{kakuno-flavortagging} 
to obtain the $b$-flavor charge $q$ and a
tagging quality factor $r \in \left[0,1\right]$.
The value $r=0$ signifies no flavor discrimination while $r=1$ implies
unambiguous flavor assignment. 
The data are divided into seven $r$ intervals.
The vertex position for the $f_{\rm rec}$ decay 
is reconstructed using the two kaon tracks from the $\phi$ meson
and that of the $f_{\rm tag}$ decay is from well-reconstructed tracks
that are not assigned to $f_{\rm rec}$~\cite{tajima-resolution}. 
The typical vertex reconstruction efficiency ($z$ resolution) is
$96\%$ ($115\,\mu$m) for $f_{\rm rec}$ and 
$94\%$ ($104\,\mu$m) for $f_{\rm tag}$.
After all selection criteria are applied, 
we obtain $75$ ($436$) events in the signal region for the $CP$ fit 
with a purity of $45\%$ ($37\%$) in the neutral (charged) mode.
%%%%%%%%%%%%%%%%%%%%%%%%%%%%%%%%%%%%
% PDFs and Resolutions Functions
%%%%%%%%%%%%%%%%%%%%%%%%%%%%%%%%%%%%
\par We determine ${\mathcal S}$ and 
${\mathcal A}$  by performing an UML fit to the observed 
$\Delta t$ distribution by maximizing the likelihood function
$\mathcal{L}({\mathcal S},{\mathcal A})=
\prod_{i} P_i({\mathcal S},{\mathcal A};\Delta t_{i})$, 
where the product is over all events in the signal region.
The likelihood $P_i$ for each event is given by
\begin{eqnarray}
P_i &=& (1-f_{\rm ol}) \int 
\biggl[\sum_j
f_j {\mathcal P}_j (\Delta t') R_j (\Delta t_i-\Delta t') 
\biggr] d(\Delta t')
\nonumber\\
&+& f_{\rm ol} P_{\rm ol}(\Delta t_i),
%\nonumber
\label{eq:likelihood}
\end{eqnarray}
where $j$ runs over signal and all background components.
${\cal P}_j(\Delta t)$ is the corresponding PDF and
$R_j(\Delta t)$ is the $\Delta t$ resolution function.
The fraction of each component ($f_j$) depends on the $r$ region and 
is calculated for each event as a function of 
$\Delta E$ and $M_{\rm bc}$.
The signal PDF
is given by a modified form of Eq.~(\ref{eq_decay}) 
by fixing $\tau_{B^0}$ and $\Delta m_d$ to their 
world-average values~\cite{pdg}
and incorporating the effect of incorrect flavor assignment.
The distribution is then convolved with a resolution function 
to take into account the finite vertex resolution.
Since the NR component is expected to have the same 
NP as the signal 
$B \to \phi K \gamma$, we treat this as signal for the
time-dependent fit~\cite{soni-fourbody}.
For the other $B\overline{B}$ components, we use the same 
functional forms as signal with an effective lifetime taken
from MC and $CP$ parameters fixed to zero.
For the continuum background, we use the functional form described
in Ref.~\cite{tajima-resolution}; 
the parameters are determined from a fit 
to the $\Delta t$ distribution of events in the data sideband 
$M_{\rm bc} < 5.26$ GeV/$c^2$ and $\Delta E \in \left[0.1,0.3\right]$ GeV.
The term $P_{\rm ol}(\Delta t)$ is a broad Gaussian function that 
represents an outlier component with a small fraction $f_{\rm ol}$.
The PDFs and resolution functions are described in detail 
elsewhere~\cite{tajima-resolution}.
%%%%%%%%%%%%%%%%%%%%%%%%%
%% Cross-checks
%%%%%%%%%%%%%%%%%%%%%%%%%
\par We perform various consistency checks of the $CP$ fitting technique.
A lifetime fit to the 
$B^0\rightarrow K^{*0}(\rightarrow K^+\pi^-)\gamma$, 
$B^+ \rightarrow \phi K^+ \gamma$ and 
$B^0 \rightarrow \phi K_S^0 \gamma$ 
data sample yields
$1.56\pm 0.03$ ps, $1.70\pm 0.20$ ps and $2.09\pm 0.45$ ps, respectively.
These are all consistent with the world-average 
values of the $B$ lifetimes.
The results of the $CP$ asymmetry fit to the
$B^0\rightarrow K^{*0}(\rightarrow K^+\pi^-)\gamma$ 
(${\mathcal S} = +0.02 \pm 0.06$, ${\mathcal A} = -0.06 \pm 0.04$)
and 
$B^+ \rightarrow \phi K^+ \gamma$ 
(${\mathcal S} = +0.25 \pm 0.33$, ${\mathcal A} = +0.18 \pm 0.26$)
are consistent with zero. 
A fit to the sideband events in the
$B^0 \to \phi K_S^0 \gamma$ data sample 
gives an asymmetry consistent with zero 
(${\mathcal S} = -1.77 \pm 1.30$, ${\mathcal A} = -0.04 \pm 0.14$).
%%%%%%%%%%%%%%%%%%%%%%%%%
%% CP fit plot
%%%%%%%%%%%%%%%%%%%%%%%%%
\begin{figure}[htbp]
  \begin{center}
    \resizebox{0.52\columnwidth}{!}{\includegraphics{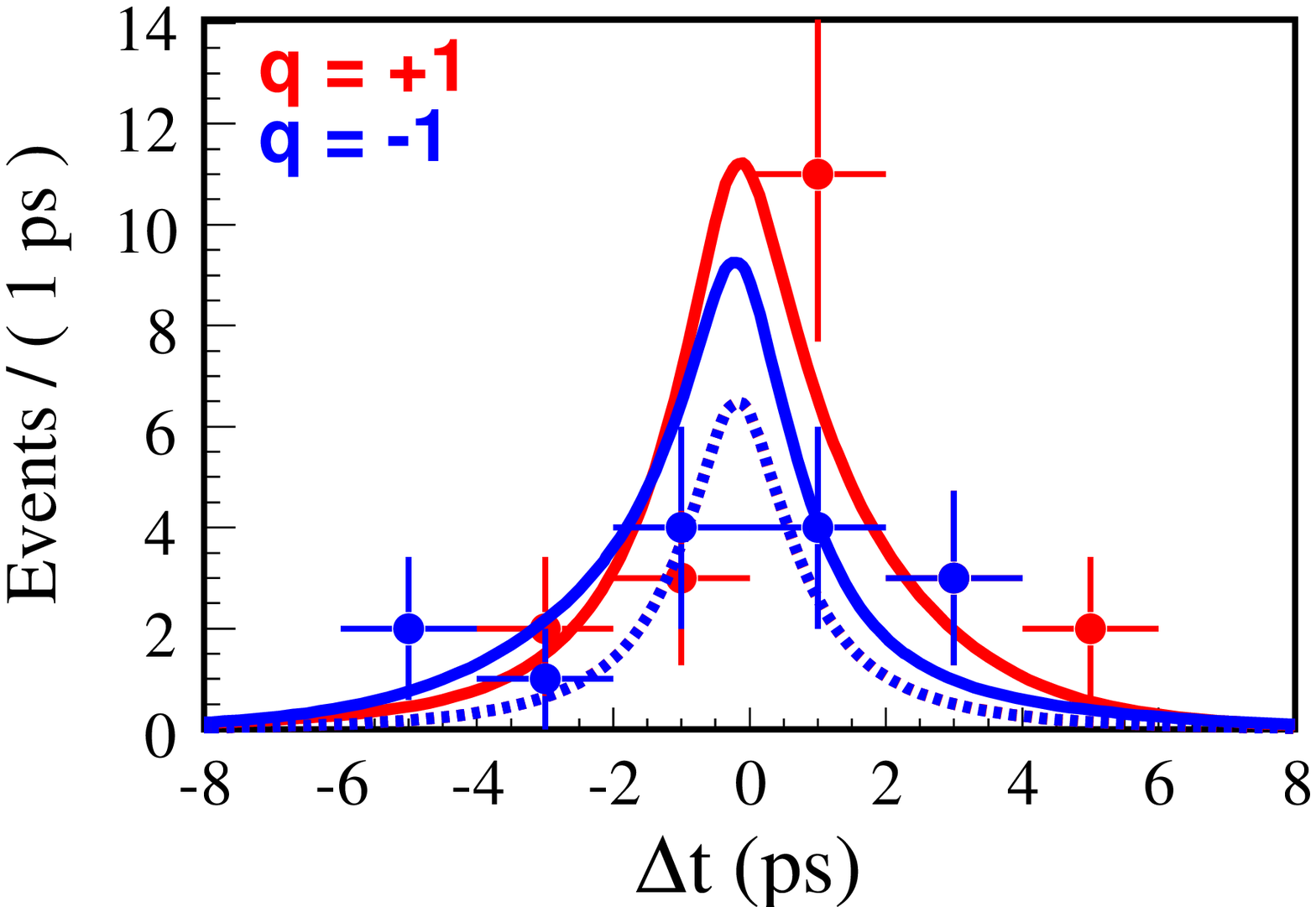}}
    \hspace{-5mm}\resizebox{0.52\columnwidth}{!}{\includegraphics{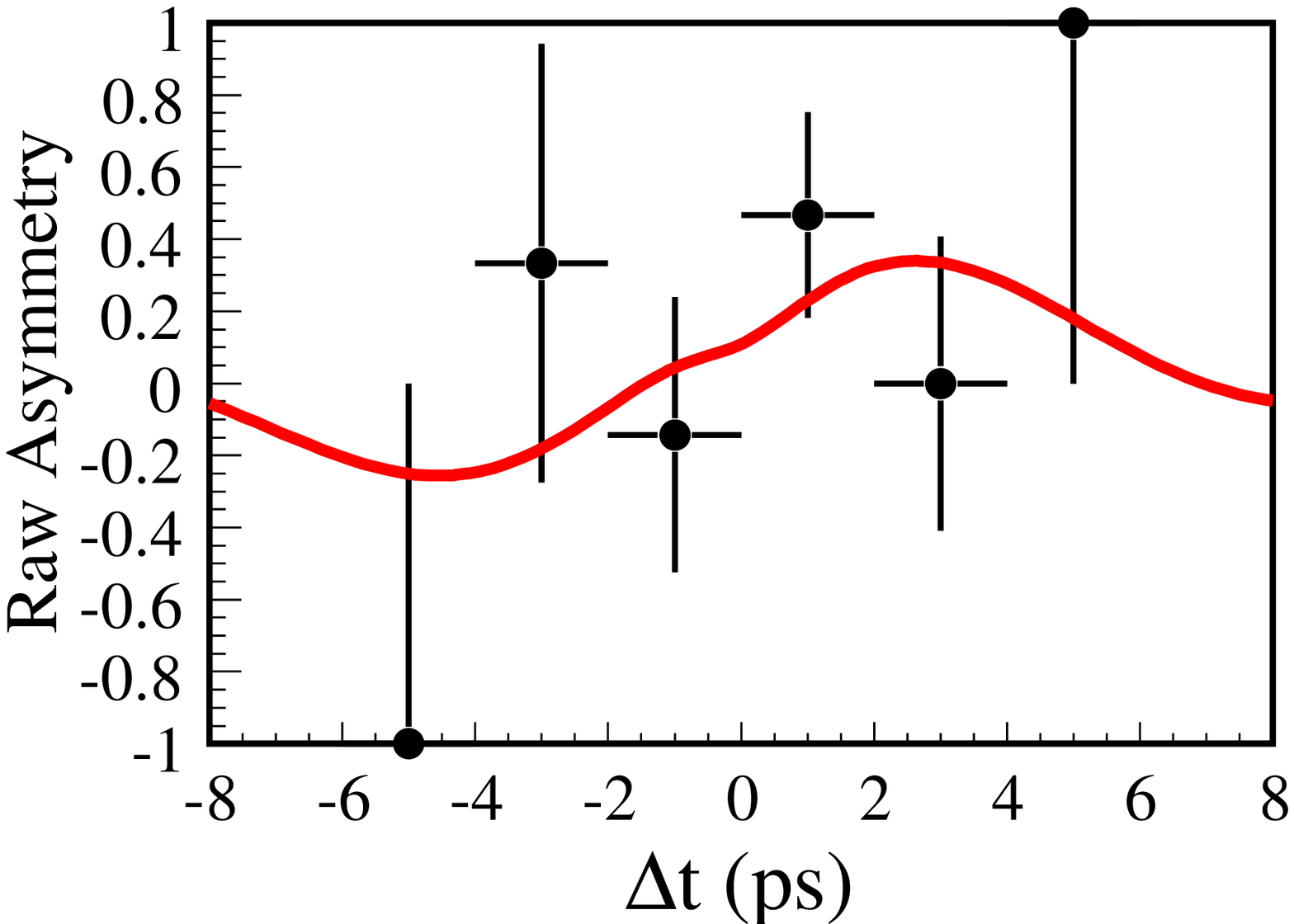}}
    \caption{$\Delta t$ distributions for $q$ = $+1$ and $q$ = $-1$ (left)
      and the raw asymmetry (right) for well-tagged events. 
      The dashed curves in the $\Delta t$ plot are the sum of 
      backgrounds while the solid curves are the sum of signal and
      backgrounds. The solid curve in the asymmetry plot shows the result
      of the UML fit.}
    \label{fig:dtandasymm}
  \end{center}
\end{figure}
%%%%%%%%%%%%%%%%%%%%%%%%%%%%%%%%%
%% CP fit fot final data sample
%%%%%%%%%%%%%%%%%%%%%%%%%%%%%%%%%
\par The only free parameters in the $CP$ fit are 
${\mathcal S}$ and ${\mathcal A}$.
The results of the fit are
${\mathcal S} = +0.74^{+0.72}_{-1.05} (\rm stat)^{+0.10}_{-0.24} (\rm syst)$ and
${\mathcal A} = +0.35 \pm 0.58 (\rm stat)^{+0.23}_{-0.10} (\rm syst)$,
where the uncertainties are obtained as described below.
We define the raw asymmetry in each $\Delta t$ bin by 
$(N_{+}-N_{-})/(N_{+}+N_{-})$, where $N_{+}$ $(N_{-})$
is the number of events with $q=+1$ $(-1)$.
Figure~\ref{fig:dtandasymm} shows the $\Delta t$ distributions 
and raw asymmetry for events with 
good tagging quality ($r > 0.5$, $48\%$ of the total).
%%%%%%%%%%%%%%%%%%%%%%%%%
%% Statistical Issues
%%%%%%%%%%%%%%%%%%%%%%%%%
\par We find that the error on ${\cal S}$ in the 
MINUIT minimization~\cite{minuitfit}
is much smaller than the expectation from MC 
simulations and has a probability of only $0.6\%$~\cite{minosvalue}.
This is due to low statistics and the presence of a single
special event (with $\Delta t = -3.64$ ps and $r = 0.96$). 
A similar effect was found in our early time-dependent
analyses of $B^0 \rightarrow \pi^+ \pi^-$~\cite{btopipi}.
Instead of the errors from MINUIT, 
we use the $\pm 68\%$ confidence intervals 
in the residual distributions of ${\cal S}$ and ${\cal A}$, 
determined from toy MC simulations
as the statistical uncertainties on the result.
%as the uncertainties on the final result.
%%%%%%%%%%%%%%%%%%%%%%%%%%%%%%%%%
%% Systematic Errors for CP fit
%%%%%%%%%%%%%%%%%%%%%%%%%%%%%%%%%
\par We evaluate the systematic uncertainties from the following sources.
A significant contribution is from the vertex 
reconstruction ($0.08$ on ${\cal S}$, $0.04$ on ${\cal A}$). 
We refit the data with each fixed parameter 
varied by its error to evaluate the uncertainties due to
signal and background fractions
($0.03$, $0.07$), 
resolution function
($0.02$, $0.03$),
$\Delta E$-$M_{\rm bc}$ shapes 
($0.01$, $0.01$),
continuum $\Delta t$ PDF
($0.01$, $0.02$),
flavor tagging
($0.01$, $0.01$)
and
effects of tagside interference~\cite{tsisyst} 
($0.004$, $0.030$).
The uncertainty from physics parameters ($\tau_{B^0}$, $\Delta m_d$),
effective lifetime and $CP$ asymmetry of the
$B\overline{B}$ background, is ($0.05$, $0.03$).
We also include 
a possible fit bias due to low statistics and 
the proximity of the central value to the physical boundary
($^{+0.00}_{-0.22}$,$^{+0.21}_{-0.00}$).
MC simulations show that this bias decreases to 0.04 with
twice the signal yield.
Adding all these contributions in quadrature, we obtain a systematic error
of $^{+0.10}_{-0.24}$ on ${\mathcal S}$ and $^{+0.23}_{-0.10}$ 
on ${\mathcal A}$.
%%%%%%%%%%%%%%%%%%%%%%%%%
%% Summary
%%%%%%%%%%%%%%%%%%%%%%%%%
\par In summary, we report the first observation of 
a new radiative decay mode, $\btopkog$
using a data sample of 
$772 \times 10^6$ $B\overline{B}$ pairs. The observed signal yield is
$37\pm8$ with a significance of $5.4\,\sigma$ 
including systematic uncertainties, and the measured branching fraction is
${\cal B}(\btopkog) = (2.74\pm 0.60 \pm 0.32) \times 10^{-6}$.
We also measure
${\cal B}(\btopkpg) = (2.48\pm 0.30 \pm 0.24) \times 10^{-6}$
with a significance of $9.6\,\sigma$.
Furthermore, we measure the charge asymmetry
${\mathcal A}_{CP} = [N(B^-)-N(B^+)]/[N(B^-)+N(B^+)] 
= -0.03\pm 0.11\pm 0.08$,
where $N(B^-)$ and $N(B^+)$ are the signal yields 
for $B^-$ and $B^+$ decays, respectively.
The signal events are mostly concentrated 
at low $\phi K$ mass near threshold.
The branching fractions and $\phi K$ mass spectra are
in agreement with the theoretical 
prediction of Ref.~\cite{hsiangnan}.
We also report the first measurements of time-dependent 
$CP$ violation parameters in the neutral mode: 
${\mathcal S} = +0.74^{+0.72+0.10}_{-1.05-0.24}$ and
${\mathcal A} = +0.35\pm 0.58^{+0.23}_{-0.10}$.
We have established that the mode 
$\btopksg$ can be used
at future high luminosity $e^+e^-$~\cite{SuperKEKB,SuperB} 
and hadronic facilities~\cite{LHCb} to perform
time-dependent $CP$ violation measurements and 
to carry out sensitive tests for NP.
%%%%%%%%%%%%%%%%%%%%%%%%%
%% Acknowledgments 
%%%%%%%%%%%%%%%%%%%%%%%%%
%***** Acknowledgments *****
%-------- Short version, if necessary, for PRL -----------
\par We thank the KEKB group for excellent operation of the
accelerator, the KEK cryogenics group for efficient solenoid
operations, and the KEK computer group and
the NII for valuable computing and SINET3 network support.  
We acknowledge support from MEXT, JSPS and Nagoya's TLPRC (Japan);
ARC and DIISR (Australia); NSFC (China); MSMT (Czechia);
DST (India); MEST, NRF, NSDC of KISTI, and WCU (Korea); MNiSW (Poland); 
MES and RFAAE (Russia); ARRS (Slovenia); SNSF (Switzerland); 
NSC and MOE (Taiwan); and DOE (USA).
%%%%%%%%%%%%%%%%%%%%%%%%%
%% References
%%%%%%%%%%%%%%%%%%%%%%%%%


\begin{thebibliography}{99}
\bibitem{conj}
Throughout this Letter, the inclusion of the charge-conjugate 
decay mode is implied unless otherwise stated.
\bibitem{charmless-pdf}
$
f(\Delta E, M_{\rm bc}) = [G_1(\Delta E - E_1) + G_2(\Delta E - E_2)] \,G(M_{\rm bc}-M_0)
$, where $G_1$, $G_2$ and $G$ are Gaussian functions and 
$E_1$, $E_2$ and $M_0$ are constants.
\bibitem{minosvalue}
The MINOS errors are $^{+0.32}_{-0.45}$ on ${\cal S}$
and $\pm 0.45$ on ${\cal A}$ for data, while the 
toy MC distributions with input values equal to those
measured in data have a width of
$^{+0.72}_{-1.05}$ for ${\cal S}$ and $\pm 0.58$ for ${\cal A}$.
The correlation between ${\cal S}$ and ${\cal A}$ 
is found to be 0.009 in data.
\end{thebibliography}
\end{document}